\documentclass[11pt,a4paper]{article}
\usepackage{graphicx}
\usepackage{amsmath,amsthm,amsfonts,amssymb,mathrsfs}
\usepackage[latin1]{inputenc}
\usepackage[english]{babel}
\newcommand{\nvec}{\nu}
\newcommand{\myvec}[1]{{\boldsymbol#1}}

\usepackage{color}

\begin{document}

\title{Design of perfectly conducting objects that are invisible to an
  incident plane wave\thanks{This work was supported by the Swedish
    Research Council under contract 2021-03720.}}

\author{Johan Helsing\thanks{Centre for Mathematical Sciences,
    Lund University Box 118, 221 00 Lund, Sweden
        (johan.helsing@math.lth.se)} \and
        Shidong Jiang\thanks{Center for Computational Mathematics,
          Flatiron Institute, Simons Foundation, New York, New York 10010
          (sjiang@flatironinstitute.org)} \and
        Anders Karlsson\thanks{Electrical and Information Technology,
    Lund University, Box 118, 221 00 Lund, Sweden (anders.karlsson@eit.lth.se)}}

%\date{(\today)}
\date{(March 4, 2024)}
\maketitle

\begin{abstract}
  This work concerns the design of perfectly conducting objects that
  are invisible to an incident transverse magnetic plane wave. The
  object in question is a finite planar waveguide with a finite
  periodic array of barriers. By optimizing this array, the amplitude
  of the scattered field is reduced to less than $10^{-9}$ times the
  amplitude of the incident plane wave everywhere outside the
  waveguide. To accurately evaluate such minute amplitudes, we employ
  a recently developed boundary integral equation technique, adapted
  for objects whose boundaries have endpoints, corners, and branch
  points.
\end{abstract}

%\allowdisplaybreaks

\section{Introduction}

This work is about the design of perfectly electrically conducting
(PEC) objects that, in a narrow frequency band, become almost
invisible to an incident plane wave in free space. The objects consist
of a finite planar waveguide, that is, two parallel PEC plates, loaded
with a finite periodic array of PEC barriers. Such barrier-loaded
(B-L) wave\-guides may serve as spatial band pass filters and they can
make other objects invisible by hiding them in their interiors. The
boundaries of the B-L wave\-guides have sharp corners, branch points
and endpoints and these {\it singular boundary points} create
challenges for numerical solvers regardless of what numerical method
they are based on. We have the ambitious criteria for invisibility
that, at the center frequency, the amplitude of the scattered field is
less than $10^{-9}$ times the amplitude of the incident field
everywhere outside the waveguide. This is to prove that the B-L
waveguides can be used as high quality spatial band pass filters. The
most common boundary integral equation (BIE) techniques, the finite
element method (FEM), and finite difference time domain techniques do
not offer the accuracy we require for geometries with singular
boundary points. For this reason we use a recently developed BIE
technique adapted for singular geometries.

A historical review of the design of invisible objects, both in
science and fiction, is given in \cite{Gbur13}. The research received
a boost in 2006, when articles on cloaking began to appear. Cloaking
means that an object is made invisible by covering it with a coating
that causes an incident wave to travel around the object such that no
scattered wave is produced. The search for materials suitable for
cloaking in turn led to intensive research in the field of
metamaterials. These materials are man-made and are designed to
possess properties not found in nature, as seen in the
reviews~\cite{Fan21,Kadic19}. Despite progress in the field of
metamaterials, a major breakthrough of cloaking technology still lies
in the future.

Another type of objects that can be made invisible are the so called
frequency selective structures (FSS) \cite{Anwar18,ShipTu12}. They
often consist of single- or multi-layer elements periodically arranged
in a plane, and can be designed to be invisible in a certain frequency
band and highly reflecting at other frequencies. The FSS are
particularly important in radar applications, but they also occur in
other applications such as dichroic sub-reflectors, microwave lenses,
and electromagnetic interference protection.

In the frequency band where the B-L waveguide is invisible, the wave
transmitted through the array of barriers has the same amplitude as
the incident plane wave. This is also the case for the transmitted
wave in certain types of microwave band-pass filters~\cite{Boria07}
and infinite or semi-infinite waveguides designed to make objects
invisible~\cite{Bonnet17,Chesnel22}. In contrast to the waveguides
in~\cite{Bonnet17} and~\cite{Boria07}, the transmitted wave of a B-L
waveguide must equal the incident wave. This condition complicates the
design of the B-L waveguide. In~\cite{Chesnel22}, an object inside a
waveguide is made invisible by deforming the walls of the
waveguide. It is an interesting approach, but it cannot be applied
here because the walls of the B-L waveguides must be planar.

Cloaked objects, FSS, and B-L waveguides can only be invisible in a
narrow frequency band. This is crucial for the B-L waveguides and for
the FSS, because frequency filtering is an important application, but
it is an undesirable feature for cloaked objects. Despite geometric
differences between the FSS and the B-L waveguides, there are
phenomenological similarities, suggesting that B-L waveguides can be
an alternative to FSS in certain applications. The B-L waveguides are
only infinite along one spatial direction, whereas the FSS are
infinite along two directions. Thus several B-L waveguides can be
combined to form large invisible structures, which, for instance, is a
desirable feature when designing radomes for radar. On the other hand,
the FSS can be made much thinner than one wavelength, while the B-L
waveguides are at least half a wavelength wide.

The paper is organized as follows. Section~\ref{sec:desc} formulates
the scattering problem and presents two types of B-L waveguides.
Section~\ref{sec:numerical} is about our BIE-based numerical scheme
for PEC objects with singular boundary points. Several designs of B-L
waveguides that are invisible in prescribed frequency bands are
presented in Section~\ref{sec:examples}. Section~\ref{sec:conclude}
contains conclusions.

\begin{figure}[t]
\centering \includegraphics[scale=0.8]{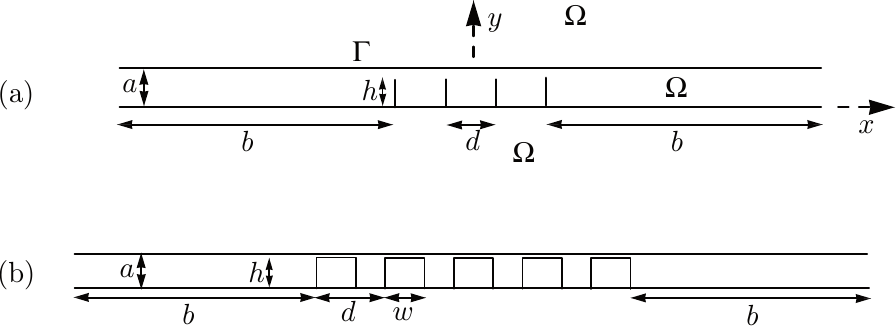}
\caption{\sf The 2D geometric cross section of two waveguides,
  translationally invariant in the $z$-direction; (a) B-L waveguide
  with four infinitely thin strips. The strips are attached to the
  lower wall. $\Gamma$ is the union of the upper and lower walls, both
  of length $2b+3d$, and the strips with total length $4h$. $\Omega$
  is the domain outside $\Gamma$ in the $xy$-plane; (b) B-L waveguide
  with five rectangular bars. The bars are attached to the lower
  wall.}
\label{geometri2}
\end{figure}

\section{Problem formulation}
\label{sec:desc}

We consider a time harmonic in-plane transverse magnetic (TM) wave
impinging on an object consisting of two disjoint PEC parts of
infinite extension in the $z$-direction, but bounded in the
$xy$-plane. The time dependence is $e^{-{\rm i}\omega t}$ and
$r=(x,y)$. The total magnetic field is
$\myvec H(r)=u(r)\hat{\myvec z}$ with decomposition
\begin{equation}
  u(r)=u^{\rm in}(x)+u^{\rm sc}(r)\,,
\label{eq:utot}
\end{equation}
where $u^{\rm in}(x)=e^{{\rm i}kx}$ is the incident field and
$u^{\rm sc}(r)$ is the scattered field. Since the waveguide is
translationally invariant in the $z$-direction, the 3D Maxwell's
equations are reduced to a scalar 2D problem in the $xy$-plane.

The boundary in the $xy$-plane is $\Gamma$. We provide $\Gamma$ with
an orientation so that an outward unit normal $\nvec=(\nu_x,\nu_y)$
can be defined. The domain outside $\Gamma$, in the $xy$-plane, is
denoted $\Omega$. In $\Omega$ there is air ($\varepsilon_{\rm r}=1$)
and the wave\-number is $k$. The side of $\Gamma$ from which $\nvec$
is directed is the positive side and the other side is the negative
side.

The scattering problem under study can be formulated as a boundary
value problem (BVP) for the Helmholtz equation
\begin{align}
  &\nabla^2 u(r)+k^2u(r)=0\,,\quad r\in\Omega\,,
                         \label{eq:PDE1}\\
  &\nvec\cdot\nabla u(r)=0\,,\quad r\in \Gamma\,,
                         \label{eq:PDE2}\\
  &u(r)=e^{{\rm i}kx}+u^{\rm sc}(r)\,,\quad r\in\Omega\,,
        \label{eq:PDE3}\\
  &u^{\rm sc}(r)=\dfrac{e^{{\rm i}k\vert r\vert}}{\sqrt{\vert r\vert}}
                 \left(F(r/\vert r\vert)+\mathcal{O}(\vert r\vert^{-1})\right),
                 \quad\vert r\vert \to \infty\,,
\label{eq:PDE4}
\end{align}
where~(\ref{eq:PDE2}) is the PEC boundary condition and
$F(r/\vert r\vert)$ is the far-field amplitude.

The fundamental solution to Helmholtz equation in the plane is
\begin{equation}
  \Phi_k(r,r')=\dfrac{\rm i}{4}H_0^{(1)}(k\vert r-r'\vert)\,,
\end{equation}
where $H_0^{(1)}$ is the zeroth-order Hankel function of the first
kind. The standard integral representation of $u^{\rm sc}(r)$ follows
from Green's theorem \cite[p.\,61]{Arfken05} and reads
\begin{equation}
  u^{\rm sc}(r)=\int_{\Gamma}v(r')
  \dfrac{\partial \Phi_k(r,r')}{\partial\nu'}\,{\rm d}\ell'\,,
  \label{intrep}
\end{equation}
where ${\rm d}\ell'$ is an element of arc length and
$\partial/\partial \nu'=\nu(r')\cdot \nabla'$. The function $v(r')$ is
given by
\begin{equation} v(r')=
\begin{cases}
  u^{\rm sc}(r')\,,\quad \text{on closed parts of $\Gamma$},\\
  u^{\rm sc+}(r')- u^{\rm sc-}(r')\,,\quad \text{on open parts of $\Gamma$},
\end{cases}
\end{equation}
where $u^{\rm sc+}(r)$ is the scattered field on the positive side of
$\Gamma$ and $u^{\rm sc-}(r)$ on the negative side. 

The invisible planar B-L waveguides designed in this work are shown in
Figure~\ref{geometri2} and they are invisible only in a single
frequency band. Their walls should be as thin as possible, and it is
reasonable to model them with zero thickness. The B-L waveguides then
have endpoints and branch points, regardless of the shape of the
barriers.

The strips in  Figure~\ref{geometri2}(a)  have zero thickness
and height $h<a$. The rectangular barriers in 
Figure~\ref{geometri2}(b)  have height $h<a$ and width $w$. The
boundary of each rectangular barrier is, thus, a closed part of
$\Gamma$.

A B-L waveguide is defined to be invisible at a wavenumber $k$ if, at
that $k$, $\vert u^{\rm sc}(r)\vert<10^{-9}$ outside the waveguide.
That is, outside the smallest rectangular domain that encloses the
waveguide. The B-L waveguides have unit inter-wall distance and are
designed to be invisible at $k$ via optimization. In the optimization
process we also require that the scattering cross section
$\sigma^{\rm sc}$, a scalar quantity defined in~(\ref{sigma10}) below,
is less than $2\pi\cdot 10^{-18}$. This bound on $\sigma^{\rm sc}$ can
be shown to hold if the bound on $\vert u^{\rm sc}(r)\vert$ holds. As
a comparison, a metal wire of unit radius has a radar cross section of
four length units in the high frequency limit.

The procedure for designing the invisible waveguides in
Figure~\ref{geometri2} is to first let $a=1$, then choose fix values
of $b$, $h$, and $w$, and finally use optimization to find $k$ and $d$
such that $\sigma^{\rm sc}<2\pi\cdot 10^{-18}$. Once $k$ and $d$ are
determined, we check that $\vert u^{\rm sc}(r)\vert<10^{-9}$ outside
the waveguide. It should be noted that one can change $k$ to a desired
value by scaling the geometry. Below, in Section~\ref{sec:relat}, we
derive an approximate relation that $k$ and $d$ must satisfy in order
for the waveguides to be invisible. This relation is used to find
initial guesses for $k$ and $d$.

The scattering cross section is defined as
\begin{equation}
\sigma^{\rm{sc}}\equiv\dfrac{\text{time average of scattered power}}
{\text{time average of incident power density}}\,.
\label{sigma10}
\end{equation}
There are several mathematically equivalent expressions for
$\sigma^{\rm sc}$. Our preferred choice for numerical evaluation in
this work is
\begin{equation}\label{sigmasc1}
  \sigma^{\rm sc}=\dfrac{1}{k}{\rm Im}\left\{\int_0^{2\pi}
    \dfrac{\partial u^{\rm sc}(r')}
    {\partial \vert r'\vert}u^{\rm sc*}(r')\vert r'\vert\,{\rm d}\phi'\right\},
\end{equation}
which follows from \eqref{sigma10} and Poynting's theorem. Here
$r'=\vert r'\vert (\cos\phi',\sin\phi')$ is a point on a circle that
circumscribes the geometric cross section of the object and the
asterisk denotes the complex conjugate.

\subsection{Waveguide modes}

To illustrate the physics of the B-L waveguides one can use the
concept of waveguide {\it modes}. Waveguide theory says that in all
sections of the wave\-guide where there are no barriers, the total
field is a sum of waveguide modes such that
\begin{equation}
  u(r)=\sum_{p=0}^\infty (\alpha_pe^{{\rm i}k_p
x}+\beta_pe^{-{\rm i}k_p x})\cos\left(p\pi y/a\right),
\end{equation}
where the wavenumber $k_p$ is given by
\begin{equation}
  k_p=\sqrt{k^2-\left(p\pi/a\right)^2},
\end{equation}
and where the amplitudes $\alpha_p$ and $\beta_p$ have different
values in all waveguide sections. The transverse electromagnetic (TEM)
mode has $p=0$. Higher-order modes have $p>0$. At the entrance of the
waveguides in Figure~\ref{geometri2} the incident TM wave couples to
the TEM mode and the higher-order modes.

A waveguide structure is invisible if the reflected TEM mode is zero
at the entrance and the transmitted TEM mode equals the incident wave
at the exit. To achieve this, the condition $k<\pi/a$ is used, which
means that all higher-order modes are evanescent. Furthermore, these
modes must decay fast enough for their amplitudes to be negligible at
the entrance and exit of the waveguide.

\subsection{A relation between the wavenumber $k$ and the period $d$} 
\label{sec:relat}

From the integral representation \eqref{intrep} it follows that the
scattered far field in the forward and backward directions are
\begin{equation}
  \label{far}
  \lim\limits_{x\rightarrow\pm \infty}u^{\rm sc}(x,0)
  =\pm\sqrt{\dfrac{k}{8\pi \vert x\vert}}e^{{\rm i}(k\vert
x\vert-\pi/4)}\int_\Gamma \nu'_xv(r')e^{\mp{\rm i} k x'}{\rm d}\ell'\,.
\end{equation}
As $\nu'_x$ is zero on the horizontal walls, only the barriers
contribute to the integral. When there is no reflected TEM mode to the
left of the first barrier there exists a wavenumber $k_{\rm s}$ for
which $\nu'_x v(r')e^{-{\rm i}k_{\rm s}x'}$ is almost the same on all
barriers. This is so because the barriers form a finite periodic
array, and it is confirmed in our numerical experiments. Thus, from
\eqref{far},
\begin{equation}
  \label{array}
  \lim\limits_{x\to \pm \infty}u^{\rm sc}(x,0)\sim
 \dfrac{1}{\sqrt{\vert x\vert}} \dfrac{1-e^{{\rm i}(k_{\rm s}\mp k)n_{\rm
bar}d}}{1-e^{{\rm i}(k_{\rm s}\mp k)d}}\,,
\end{equation}
where $n_{\rm bar}$ is the number of barriers.

The two-dimensional optical theorem~\cite[eq.~(39)]{HelsKarl13}
\begin{equation}
  \label{cross}
  \sigma^{\rm sc}=-\lim\limits_{x\rightarrow\infty}\sqrt{\dfrac{8\pi x}{k}}
  {\rm Re}\left\{u^{\rm sc}(x,0)e^{-{\rm i}(k x-\pi/4)}\right\}
\end{equation}
states that a necessary and sufficient condition for
$\sigma^{\rm{sc}}=0$ is
$\lim\limits_{x\to \infty}\sqrt{x}u^{\rm sc}(x,0)=0$. Another
necessary condition for invisibility is
$\lim\limits_{x\to -\infty}\sqrt{\vert x\vert}u^{\rm sc}(x,0)=0$, as
seen from \eqref{sigmasc1}. A relation between $k$ and $d$, and also a
relation between $k_{\rm s}$ and $d$, now follows by applying
$\lim\limits_{x\to \pm \infty}\sqrt{\vert x\vert}u^{\rm sc}(x,0)=0$ to
\eqref{array}:
\begin{align}
  &kn_{\rm bar}d\approx n\pi
\label{constraint2a}\,,\\
  &k_{\rm s}n_{\rm bar}d \approx n\pi+2m\pi\,,
\label{constraint2b}
\end{align}
where $n,m>0$ are integers. Although approximate, \eqref{constraint2a}
holds to at least three digits for the invisible waveguides that we
have designed in this work.

\section{Numerical scheme}
\label{sec:numerical}

In this section we first derive the BIE reformulation of the
BVP~(\ref{eq:PDE1}), (\ref{eq:PDE2}), (\ref{eq:PDE3}), and
(\ref{eq:PDE4}), upon which our numerical scheme is based. Then we
discuss how to efficiently discretize and solve this BIE in the
presence of singular boundary points and how to subsequently evaluate
the solution, $u(r)$ or $u^{\rm sc}(r)$, at points $r$ in the
computational domain. We end with some theoretical considerations
regarding the second-kindness of our BIE.

\subsection{Integral equation reformulation}

Let, as in Section~\ref{sec:desc}, $\Gamma$ be a union of oriented
smooth open arcs, referred to collectively as the {\it boundary}. The
outward unit normal at $r\in\Gamma$ is $\nu(r)$. We use standard
definitions of the acoustic layer operators $S_k$, $K_k$, and
$T_k$~\cite[pp.~41--42]{ColtonKress98}, defined by their action on a
layer density $\rho(r)$ on $\Gamma$ as
\begin{align}
S_k\rho(r)&=2\int_{\Gamma}\Phi_k(r,r')\rho(r')\,{\rm d}\ell'\,, 
\quad r\in\Gamma\,,
\label{eq:Soper}\\
K_k\rho(r)&=
  2\int_{\Gamma}\frac{\partial\Phi_k}{\partial\nu'}(r,r')\rho(r')
            \,{\rm d}\ell'\,,\quad r\in\Gamma\,,
\label{eq:Koper}\\
T_k\rho(r)&=
  2\int_{\Gamma}\frac{\partial^2\Phi_k}
  {\partial\nu\partial\nu'}(r,r')\rho(r')\,{\rm d}\ell'\,,
\quad r\in\Gamma\,,
\label{eq:Toper}
\end{align}
where $\partial/\partial\nu=\nu(r)\cdot\nabla$. For ease of
presentation we also use $K_k$ to define the double-layer
potential. That is, we use
\begin{equation}
K_k\rho(r)=
  2\int_{\Gamma}\frac{\partial\Phi_k}{\partial\nu'}(r,r')\rho(r')
  \,{\rm d}\ell'\,,\quad r\in\mathbb{R}^2\setminus\Gamma\,.
\end{equation}

Within our BIE framework, the field $u^{\rm sc}(r)$ of the Neumann
Helmholtz problem~(\ref{eq:PDE1}), (\ref{eq:PDE2}), (\ref{eq:PDE3}),
and (\ref{eq:PDE4}) has the representation
\begin{equation}
  u^{\rm sc}(r)=K_k(-S_k)\rho(r)\,,\quad r\in\mathbb{R}^2\setminus\Gamma\,.
\label{eq:urepN}
\end{equation}
This representation is recommended for piecewise smooth open arcs
$\Gamma$ in~\cite{HelsJian23} and also recommended, along with
singular weight functions, for everywhere smooth open arcs $\Gamma$
in~\cite{BrunLint12}. The BIE itself, obtained by
inserting~(\ref{eq:PDE3}) with~(\ref{eq:urepN}) in~(\ref{eq:PDE2}), is
\begin{equation}
  T_k(-S_k)\rho(r)=-\nu(r)\cdot\nabla e^{{\rm i}kx}\,,\quad r\in\Gamma\,.
\label{eq:Neum}
\end{equation}

We plan to solve~(\ref{eq:Neum}) for $\rho(r)$ using Nyström
discretization accelerated by the recursively compressed inverse
preconditioning (RCIP) method. While there is nothing new with using
Nyström/RCIP schemes for solving BIEs per se, see~\cite{Tutorial} and
references therein, it may not be obvious which is the best way to
apply such schemes to BIEs with composed singular operators as
in~(\ref{eq:Neum}). We therefore now devote Section~\ref{sec:overview}
to an overview of this topic.

\subsection{Overview of RCIP}
\label{sec:overview}

The RCIP method accelerates and stabilizes Nyström solvers in certain
``singular'' situations. More precisely, RCIP applies to the Nyström
discretization of Fredholm second kind BIEs on closed or open $\Gamma$
that contain some type of singular boundary points. Our overview,
based on~\cite[Sections~3 and~16]{Tutorial}, summarizes the most
important features of RCIP with particular emphasis on composed
operators. Section~\ref{sec:standard} reviews RCIP in a standard
setting and Section~\ref{sec:compose} shows what modifications are
needed for~(\ref{eq:Neum}). The Nyström discretization is assumed to
rely on composite Gauss--Legendre quadrature with $16$ nodes per
quadrature panel. See~\cite{HelsJian23} for more detail.

\subsubsection{A Fredholm integral equation in standard form}
\label{sec:standard}

We start with a Fredholm second kind BIE on $\Gamma$ in standard form
\begin{equation}
\left(I+G\right)\rho(r)=g(r)\,,\quad r\in \Gamma\,.
\label{eq:1}
\end{equation}
Here $I$ is the identity, $G$ is an integral operator that is often
assumed compact away from singular boundary points on $\Gamma$, $g(r)$
is a piecewise smooth right-hand side, and $\rho(r)$ is the unknown
layer density to be solved for.

For simplicity, we only discuss how to deal with one singular point on
$\Gamma$ and denote this point $\gamma$. Multiple singular points can
be treated independently and in parallel.

Let $G$ be split into two parts
\begin{equation}
G=G^\star+G^\circ\,,
\label{eq:splitG}
\end{equation}
where $G^\star$ describes the kernel interaction close to $\gamma$ and
$G^\circ$ is the (compact) remainder. Let the part of $\Gamma$ on
which $G^\star$ is nonzero be denoted $\Gamma^\star$.

Now introduce the {\it transformed density}
\begin{equation}
\tilde{\rho}(r)=\left(I+G^\star\right)\rho(r)\,.
\label{eq:tilde}
\end{equation}
Then use~(\ref{eq:splitG}) and~(\ref{eq:tilde}) to
rewrite~(\ref{eq:1}) as
\begin{equation} \left(I+G^\circ(I+G^\star)^{-1}\right)\tilde{\rho}(r)
=g(r)\,,\quad r\in \Gamma\,.
\label{eq:2}
\end{equation}
Although~(\ref{eq:2}) looks similar to~(\ref{eq:1}), there are
advantages to using~(\ref{eq:2}) rather than~(\ref{eq:1}) from a
numerical point of view: the spectral properties of
$G^\circ(I+G^\star)^{-1}$ are better than those of $G$.

The Nyström/RCIP scheme discretizes~(\ref{eq:2}) chiefly on a grid on
a {\it coarse mesh} of panels on $\Gamma$ that is sufficient to
resolve $G^\circ$ and $g(r)$. Only the inverse $(I+G^\star)^{-1}$
needs a grid on a local, dyadically refined, {\it fine mesh} of panels
on $\Gamma^\star$. The discretization of~(\ref{eq:2}) takes the form
\begin{equation}
  \left({\bf I}_{\rm coa}+{\bf G}_{\rm coa}^\circ{\bf
      R}\right) \tilde{\boldsymbol{\rho}}_{\rm coa}={\bf g}_{\rm coa}\,,
\label{eq:3}
\end{equation}
where ${\bf R}$ is a sparse block matrix called the {\it compressed
  inverse} and where the discrete unknown
$\tilde{\boldsymbol{\rho}}_{\rm coa}$ lives on the coarse grid only.
The boundary part $\Gamma^\star$ contains the two, four, or six coarse
panels closest to $\gamma$ depending on if $\gamma$ is an endpoint, a
corner, or a branch point.

The power of RCIP lies in the construction of ${\bf R}$. In theory,
${\bf R}$ corresponds to a discretization of $(I+G^\star)^{-1}$ on the
fine grid, followed by a lossless compression to the coarse grid. In
practice, ${\bf R}$ is constructed via a forward recursion on a
hierarchy of grids on $\Gamma^\star$ and where refinement and
compression occur in tandem. The computational cost grows, at most,
linearly with the number of refinement levels.

Once the compressed equation~(\ref{eq:3}) is solved for
$\tilde{\boldsymbol{\rho}}_{\rm coa}$ and its {\it weight-corrected}
counterpart
$\hat{\boldsymbol{\rho}}_{\rm coa}={\bf
  R}\tilde{\boldsymbol{\rho}}_{\rm coa}$ is produced, several useful
functionals of $\rho(r)$ can be computed with ease. Should one so
wish, the solution $\rho(r)$ on the fine grid can be reconstructed
from $\tilde{\boldsymbol{\rho}}_{\rm coa}$ via a backward recursion on
the same hierarchy of grids that is used in the construction of
${\bf R}$.

\subsubsection{An integral equation with composed operators}
\label{sec:compose}

The BIE~(\ref{eq:Neum}), that needs to be solved in the present work,
is cast in the form
\begin{equation}
  A(-B)\rho_1(r)=g(r)\,,\quad r\in \Gamma\,,
\label{eq:AB}
\end{equation}
where the unknown density now is denoted $\rho_1(r)$ and where
$A=T_k$, $B=S_k$, and $g(r)=-\nu(r)\cdot\nabla e^{{\rm i}kx}$.
Clearly, the composition of $A$ and $-B$ and the absence of $I$
make~(\ref{eq:AB}) look quite different from the standard
form~(\ref{eq:1}). It should be noted, however, that for versions of
$A=T_k$ and $B=S_k$, weighted as in Chebyshev--Gauss quadrature, and
on smooth open $\Gamma$, the BIE~(\ref{eq:AB}) is mathematically
equivalent to a BIE in the form~(\ref{eq:1}) with $G$ being a compact
operator~\cite[Theorem~1]{BrunLint12}.

We now apply the Nyström/RCIP scheme of Section~\ref{sec:standard} to
(\ref{eq:AB}), with the goal to find the analogue of the compressed
discrete equation~(\ref{eq:3}) for~(\ref{eq:AB}). First, introduce the
new temporary density $\rho_2(r)$ via
\begin{equation}
\rho_2(r)=-B\rho_1(r)\,.
\end{equation}
This allows us to rewrite~(\ref{eq:AB}) as the $2\times 2$ block
system
\begin{equation}
\left(
\begin{bmatrix}
I & 0 \\
0 & I
\end{bmatrix} 
+
\begin{bmatrix}
 -I & A \\
  B & 0
\end{bmatrix} 
\right)
\begin{bmatrix}
 \rho_1(r) \\ 
 \rho_2(r)
\end{bmatrix}
=
\begin{bmatrix}
   g(r) \\ 0 
\end{bmatrix}\,,
\label{eq:11}
\end{equation}
which corresponds to~(\ref{eq:1}) in Section~\ref{sec:standard}.

Note that~(\ref{eq:11}) is free from composed operators so that the
Nyström/RCIP scheme can be applied. The analogue of~(\ref{eq:2}) is
\begin{equation}
\left(
\begin{bmatrix}
I & 0 \\
0 & I
\end{bmatrix} 
+
\begin{bmatrix}
 -I^\circ & A^\circ \\
  B^\circ & 0
\end{bmatrix} 
\left(
\begin{bmatrix}
I & 0 \\
0 & I
\end{bmatrix} 
+
\begin{bmatrix}
 -I^\star & A^\star \\
  B^\star & 0
\end{bmatrix} 
\right)^{-1}\right)
\begin{bmatrix}
 \tilde{\rho}_1(r) \\ 
 \tilde{\rho}_2(r)
\end{bmatrix}
=
\begin{bmatrix}
   g(r) \\ 0 
\end{bmatrix}.
\label{eq:22}
\end{equation}
The analogue of~(\ref{eq:3}) is
\begin{equation}
\left(
\begin{bmatrix}
{\bf I}_{\rm coa} & {\bf 0}_{\rm coa} \\
{\bf 0}_{\rm coa} & {\bf I}_{\rm coa}
\end{bmatrix} 
+
\begin{bmatrix}
 -{\bf I}^\circ_{\rm coa} & {\bf A}^\circ_{\rm coa} \\
  {\bf B}^\circ_{\rm coa} & {\bf 0}_{\rm coa}
\end{bmatrix} 
\begin{bmatrix}
{\bf R}_1 & {\bf R}_3 \\
{\bf R}_2 & {\bf R}_4
\end{bmatrix} 
\right)
\begin{bmatrix}
 \tilde{\boldsymbol{\rho}}_{1{\rm coa}} \\ 
 \tilde{\boldsymbol{\rho}}_{2{\rm coa}}
\end{bmatrix}
=
\begin{bmatrix}
   {\bf g}_{\rm coa} \\ 0 
\end{bmatrix}.
\label{eq:33}
\end{equation}
Here we have partitioned the compressed inverse ${\bf R}$ into the
four sparse equi-sized blocks ${\bf R}_1$, ${\bf R}_2$ ${\bf R}_3$,
and ${\bf R}_4={\bf I}^\circ_{\rm coa}$.

As an extra twist, we write the compressed equation~(\ref{eq:33}) as a
linear system involving only one discrete density -- not two. To this
end, introduce the new density $\tilde{\boldsymbol{\rho}}_{\rm coa}$
via
\begin{align}
  \tilde{\boldsymbol{\rho}}_{1{\rm coa}}&=
  \tilde{\boldsymbol{\rho}}_{\rm coa}+
  {\bf R}_1^{-1}{\bf R}_3{\bf B}_{\rm coa}^\circ{\bf R}_1
  \tilde{\boldsymbol{\rho}}_{\rm coa}\,,
\label{eq:single1}\\
  \tilde{\boldsymbol{\rho}}_{2{\rm coa}}&=
  -{\bf B}_{\rm coa}^\circ{\bf R}_1
  \tilde{\boldsymbol{\rho}}_{\rm coa}\,.
\label{eq:single2}
\end{align}
The change of variables~(\ref{eq:single1}) and (\ref{eq:single2}) is
chosen so that the second block-row of~(\ref{eq:33}) is automatically
satisfied. The first block-row of~(\ref{eq:33}) becomes
\begin{multline}
  \left[{\bf I}_{\rm coa}^\star
    -{\bf A}_{\rm coa}^\circ
\left({\bf R}_4-{\bf R}_2{\bf R}_1^{-1}{\bf R}_3\right)
{\bf B}_{\rm coa}^\circ{\bf R}_1+{\bf A}_{\rm coa}^\circ{\bf R}_2
\right.\\
\left.+{\bf R}_1^{-1}{\bf R}_3{\bf B}_{\rm coa}^\circ{\bf R}_1
\right]\tilde{\boldsymbol{\rho}}_{\rm coa}={\bf g}_{\rm coa}\,.
\label{eq:disc3}
\end{multline}

We observe, from~(\ref{eq:single1}), that away from $\Gamma^\star$ the
discrete density $\tilde{\boldsymbol{\rho}}_{\rm coa}$ coincides with
$\tilde{\boldsymbol{\rho}}_{1{\rm coa}}$. Therefore one can expect the
system~(\ref{eq:disc3}) to share more properties with the original
equation~(\ref{eq:AB}) than the expanded system~(\ref{eq:33}).
Numerical experiments indicate that~(\ref{eq:disc3}) is superior
to~(\ref{eq:33}) not only in terms of computational economy, but also
in terms of stability and of convergence of iterative solvers.

\subsection{On-surface and near-field evaluation}

The discretization of the BIE~(\ref{eq:AB}) on $\Gamma$ involves the
discretization of integrals with various types of singular
integrands. Furthermore, for $r$ close to $\Gamma$, the discretization
of the field representation~(\ref{eq:urepN}) involves various types of
nearly singular integrands. These singular and nearly singular
integrals can, typically, not be accurately evaluated at a reasonable
cost using composite standard interpolatory quadrature, but require
special discretization techniques. For this, we use a panel-based
product integration scheme which, according to the classification
of~\cite{HaBaMaYo14}, is ``explicit split''.

Our panel-based explicit-split quadrature scheme is described in
detail in~\cite[Section~4]{HelsKarl18}. For now, we only mention that
the scheme combines recursion with analytical methods such as the use
of the fundamental theorem of complex line integrals. Vectors $r$ and
$r'$ in the real plane $\mathbb{R}^2$ are identified with points $z$
and $\tau$ in the complex plane $\mathbb{C}$. Nearly singular and
singular integral operator kernels $G(r,r')$ are split according to
the general pattern
\begin{multline}
G(r,r')\,{\rm d}\ell'=G_0(r,r')\,{\rm d}\ell'
+\log|r-r'|G_{\rm L}(r,r')\,{\rm d}\ell'\\
+c_{\rm C}{\rm Re}\left\{\frac{G_{\rm C}(z,\tau)\,{\rm d}\tau}
                 {{\rm i}(\tau-z)}\right\}
+{\rm Re}\left\{\frac{G_{\rm H}(z,\tau)\,{\rm d}\tau}
                 {{\rm i}(\tau-z)^2}\right\}\,.
\label{eq:ex3}
\end{multline}
Here $c_{\rm C}$ is a, possibly complex, constant and $G_0(r,r')$,
$G_{\rm L}(r,r')$, $G_{\rm C}(z,\tau)$, and $G_{\rm H}(z,\tau)$ are
smooth functions. Our scheme requires explicit formulas for $G(r,r')$,
$G_{\rm L}(r,r')$, $G_{\rm C}(z,\tau)$, and $G_{\rm H}(z,\tau)$, while
$G_0(r,r')$ needs only to be known in the limit $r'\to
r$. See~\cite[Section~6.2]{Ande23} for additional remarks.

\subsection{Second-kindness and compact operators}

The assumption, mentioned in Section~\ref{sec:standard}, that
$G^\circ$ of~(\ref{eq:splitG}) should be compact for RCIP to apply, is
actually too restrictive. For example, in~\cite{HelsOjal09} it is
demonstrated that $G^\circ$ can contain a Cauchy-singular operator:
using the GMRES iterative solver in the solve phase, RCIP still works
although the number of GMRES iterations needed for convergence may be
high due to the presence of multiple accumulation points in the
spectrum of $G^\circ$. See~\cite[Eq.~(24)]{Tutorial} for a better,
although more technical, assumption about $G^\circ$.

Allowing $G^\circ$ of~(\ref{eq:splitG}) to be non-compact is important
in the present work. In~(\ref{eq:AB}) we view $B$ as a reasonable
right preconditioner to $A$ and merely consider $(A(-B)-I)^\circ$ as
an operator whose spectrum is clustered.

\begin{figure}[t]
  \centering
  \includegraphics[height=46mm]{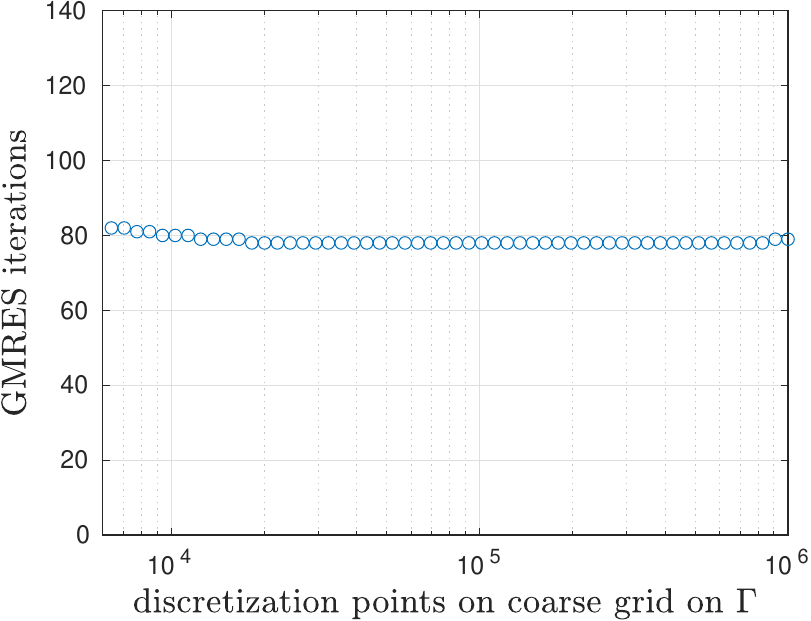}
  \caption{\sf Behavior of~(\ref{eq:AB}) with~(\ref{eq:disc3}) under
    uniform overresolution for the Neumann Helmholtz problem of
    scattering of a TM plane wave by the spiral-shaped arc
    of~\cite[Figure~2]{BrunLint12} and with $k=104.8530264831035$: GMRES
    iterations for full convergence.}
  \label{fig:overresspiral}
\end{figure}

Some light can be shed on the ``second-kindness'' of (\ref{eq:disc3})
on different $\Gamma$ by monitoring the convergence of GMRES under
uniform overresolution. For example, Figure~\ref{fig:overresspiral}
shows that on a smooth open $\Gamma$, the number of GMRES iterations
required to meet a small stopping criterion threshold in the relative
residual can be bounded -- indicating that~(\ref{eq:disc3}) behaves as
a discretization of a BIE in the form~(\ref{eq:1}) with compact $G$.

A related issue is what the authors of~\cite{AgaOneRac23} call
``numerical second-kindness'': even if a BIE in the form~(\ref{eq:AB})
is mathematically equivalent to a second kind BIE~(\ref{eq:1}) with
compact $G=A(-B)-I$, it may happen that the GMRES iterations
stall. The reason being that $A(-B)-I$ is not available to as high
precision as $A$ and $B$ themselves due to numerical
cancellation. Then the formulation is said to lack numerical
second-kindness~\cite[Section~6.1]{AgaOneRac23}. In the present work
we have not observed any such bad behavior.

\section{Examples of B-L waveguides}
\label{sec:examples}

We now present numerical results for two types of B-L waveguides. The
numerical codes are written in {\sc Matlab}, release 2022a, and
executed on a workstation equipped with an Intel Core i7-3930K CPU and
$64$ GB of RAM. The built-in function {\tt fminsearch} is used for
design optimization. The fast multipole method routine {\tt hfmm2d}
from~\cite{fmm2d} is used in connection with matrix-vector
multiplication whenever deemed efficient.

When assessing the accuracy of computed fields and scattering cross
sections we adopt a procedure where to each numerical solution we also
compute an overresolved reference solution, using roughly $50\%$ more
points in the discretization of the setup under study and doubled
radius $\vert r'\vert$ in~(\ref{sigmasc1}). The absolute difference
between the two solutions at a field point is the {\it estimated
  absolute pointwise error}. The fields and their errors are always
computed at $10^6$ field points on a rectangular Cartesian grid in the
computational domains shown.

\subsection{B-L waveguide with strip barriers}

Consider first the design of a B-L waveguide with vertical strips, as
in Figure~\ref{geometri2}(a). The distance between the walls, $a$, is
unity in a suitable unit of length, as stated in
Section~\ref{sec:desc}. For waveguides with more than three strips,
there is a range for each of $b$, $d$, $h$, and $k$ in which one can
find values that give $\sigma^{\rm{ sc}}<2\pi\cdot 10^{-18}$.  In the
examples shown here, values of $b$ and $h$ are first chosen in these
ranges. The values of $k$ and $d$ are then determined by {\tt
  fminsearch} via minimization of $\sigma^{\rm{ sc}}$.

\begin{figure}[t]
  \centering
  \includegraphics[height=47mm]{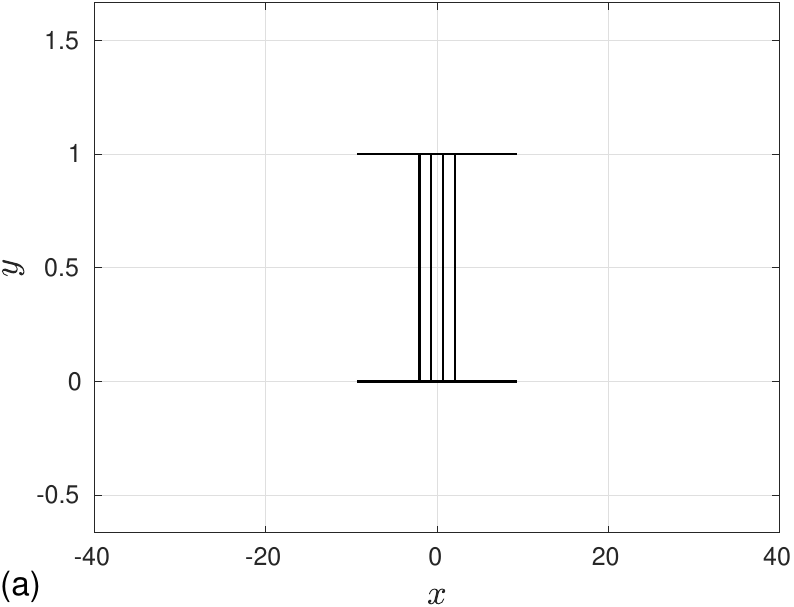}
  \includegraphics[height=47mm]{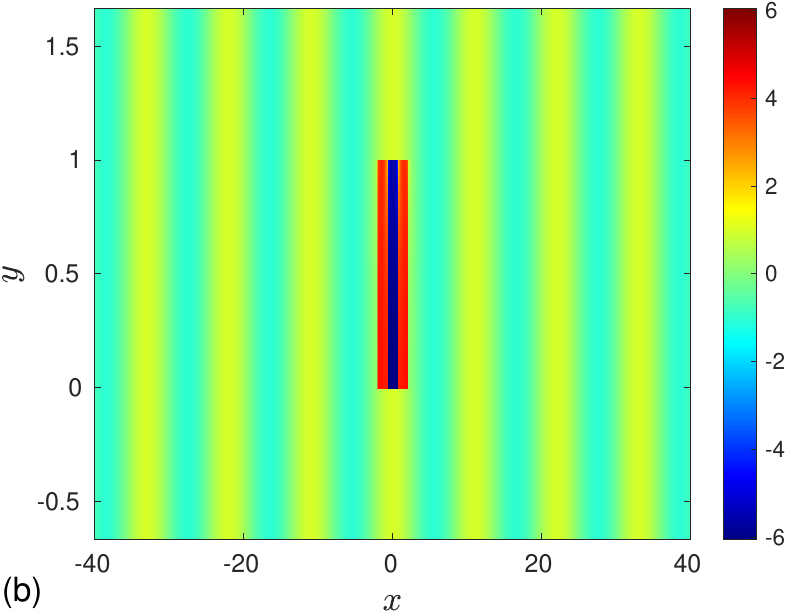}
  \includegraphics[height=47mm]{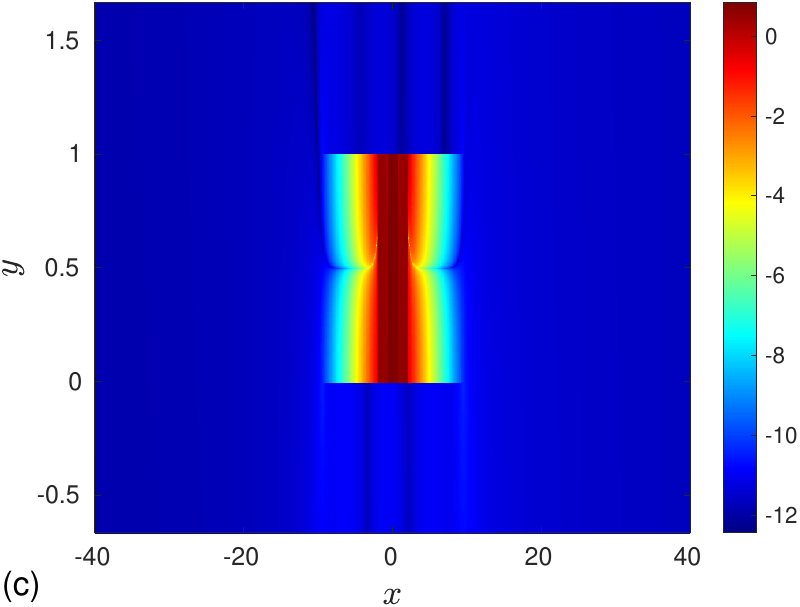}
  \includegraphics[height=47mm]{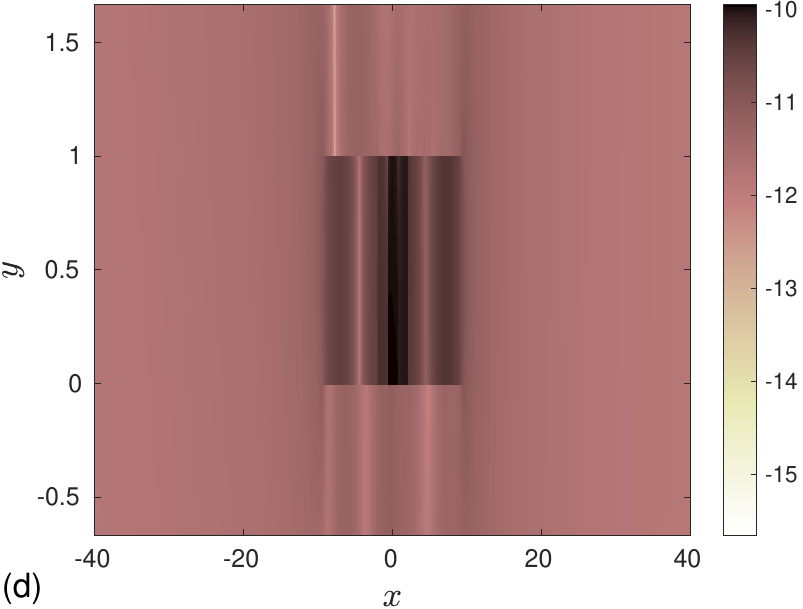}
  \caption{\sf (a) geometry of the four-strip waveguide; (b) real part
    of $u(r)$; (c) $\log_{10}$ of $\vert u^{\rm sc}(r)\vert$; (d)
    $\log_{10}$ of estimated absolute pointwise error in
    $u^{\rm sc}(r)$.}
\label{konfig8d}
\end{figure}
  
\begin{figure}[t]
  \centering \includegraphics[height=46mm]{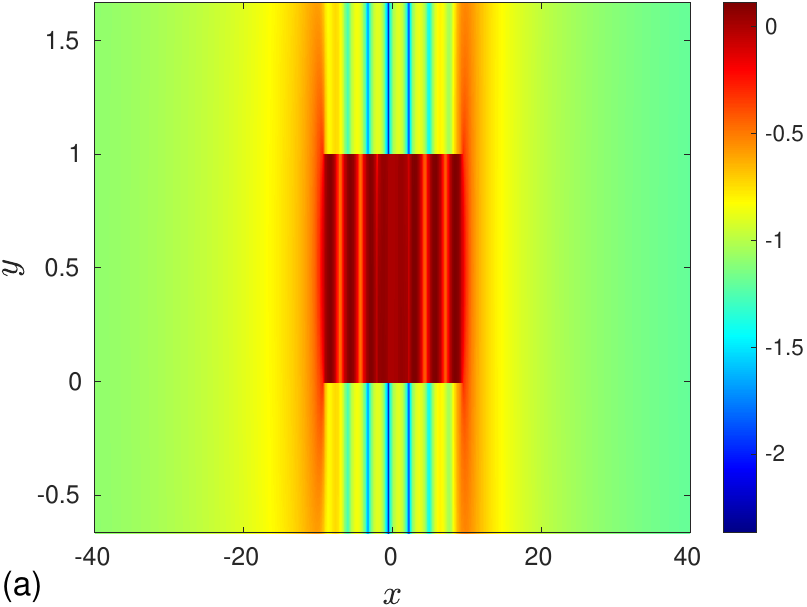}
\includegraphics[height=46mm]{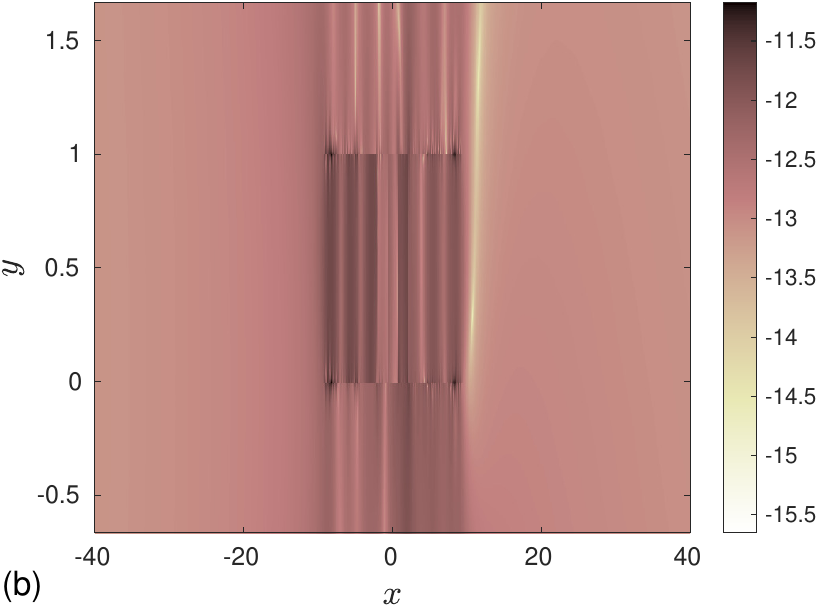}
\caption{\sf The four-strip waveguide at doubled wavenumber: (a)
$\log_{10}$ of $\vert u^{\rm sc}(r)\vert$; (b) $\log_{10}$ of
estimated absolute pointwise error in $u^{\rm sc}(r)$.}
\label{konfig8d2k}
\end{figure}

Figures~\ref{konfig8d}, \ref{konfig8d2k}, \ref{konfig14a}, and
\ref{konfig8d8p} show fields and error estimates computed at field
points in the computational domain $\{-40\le x\le 40, -2/3\le y\le
5/3\}$. The $x$- and $y$-axes are scaled differently to provide better
field resolution inside the waveguide, given the aspect ratio of the
images and the range of $x$. The radius in~(\ref{sigmasc1}) is
chosen as $\vert r'\vert=30$.

\subsubsection{B-L waveguide with four strips}
\label{sec:fourstrip}

The first example is a four-strip waveguide with $a=1$, $b=7.3$ and
$h=0.9973$. Optimization gives
$\sigma^{\rm{ sc}}\approx 2\cdot 10^{-21}$ at $k=0.5712887729818$ and
$d=1.373106370502$. This $\sigma^{\rm{ sc}}$ is small enough to be at
the overall noise level of the numerical scheme.
Figure~\ref{konfig8d}(a) shows the waveguide. The gaps between the
endpoints of the strips and the upper wall are too small to be
seen. Figure~\ref{konfig8d}(b) shows the real part of $u(r)$. Only in
the regions between the strips is there a noticeable difference
between $u(r)$ and $u^{\rm in}(x)$. The field pattern shows that $n=1$
and $m=1$ in \eqref{constraint2a} and \eqref{constraint2b}. The
$\log_{10}$ of $\vert u^{\rm sc}(r)\vert$ is shown in
Figure~\ref{konfig8d}(c). From the first strip to the waveguide
entrance, and from the last strip to the waveguide exit, the scattered
field is attenuated due to evanescent modes. The $\log_{10}$ of
estimated absolute pointwise error in the scattered field is shown in
Figure~\ref{konfig8d}(d). A comparison with Figure~\ref{konfig8d}(c)
reveals that the amplitude of $u^{\rm sc}(r)$ is down to the numerical
noise level outside the waveguide.
 
As for computational performance we quote the following: with the
given values of $a$, $b$, $d$, $h$, and $k$, the setup in
Figure~\ref{konfig8d} uses $5,\!344$ discretization points on $\Gamma$
and $83$ iterations in GMRES to meet a stopping criterion threshold of
$10^{-14}$ in the estimated relative residual for the main linear
system. The total execution time, including field evaluation at the
$10^6$ field points, is around $25$ seconds.

\subsubsection{B-L waveguide with four strips at doubled wavenumber}

From (\ref{constraint2a}) and (\ref{constraint2b}) one might get the
impression that if a B-L waveguide is invisible at $k$, then it is
also invisible at wave numbers close to $Nk$ for integers $N>0$. This
is not true since (\ref{constraint2a}) and (\ref{constraint2b}) are
necessary but not sufficient conditions for invisibility. When $k$
from Section~\ref{sec:fourstrip} is doubled while $b$, $d$, and $h$
are kept the same, then $\sigma^{\rm sc}=9.1\cdot 10^{-1}$. Also the
scattered near field becomes large, as seen in
Figure~\ref{konfig8d2k}.

\subsubsection{Influence of length of waveguide}

The B-L waveguides must have $b$ large enough for the evanescent
TM-modes to become negligible at the entrance and at the exit. In
Figure~\ref{konfig8d}(a) the sufficient value $b=7.3$ is used. Smaller
values of $b$ affect the minimum value of $\sigma^{\rm sc}$ that can
be reached: $b=5$ gives $\sigma^{\rm sc}=2.4\cdot 10^{-15}$ and
$b=4.3$ gives $\sigma^{\rm sc}=1.6\cdot 10^{-13}$.

\begin{figure}[t]
  \centering
  \includegraphics[height=47mm]{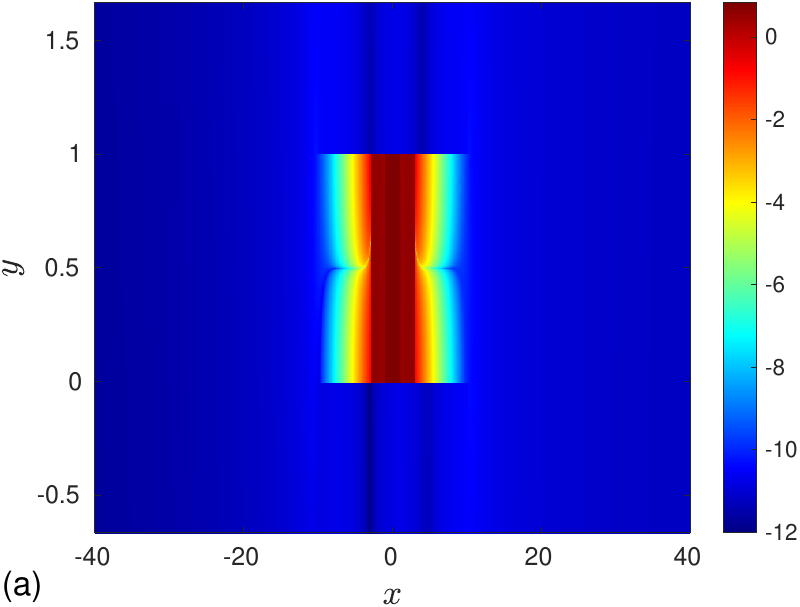}
  \includegraphics[height=47mm]{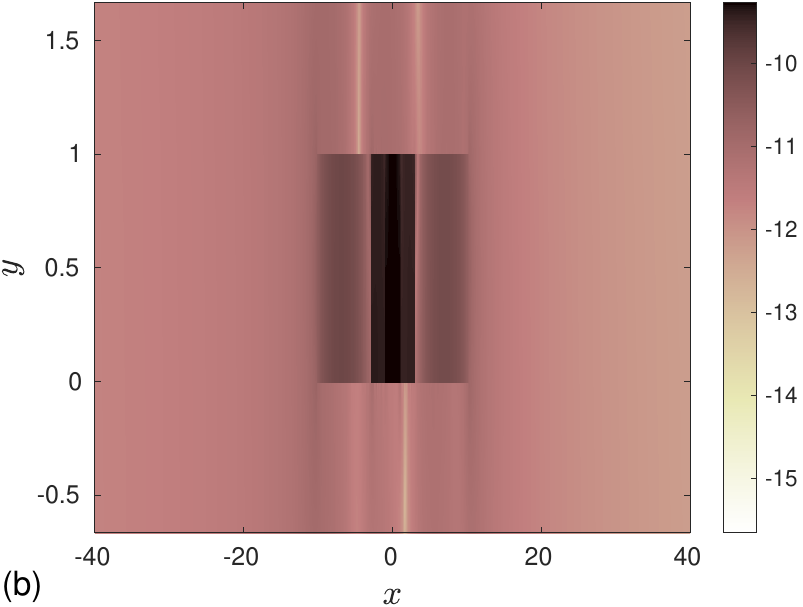}
  \caption{\sf The four-strip waveguide with $h=0.9997$: (a)
    $\log_{10}$ of $\vert u^{\rm sc}(r)\vert$; (b) $\log_{10}$ of
    estimated absolute pointwise error in $u^{\rm sc}(r)$.}
\label{konfig14a}
\end{figure}

\subsubsection{B-L waveguide with four strips and smaller gaps}

The gap $a-h$, between the upper wall and the endpoints of the strips,
can be both smaller and larger than in Figure~\ref{konfig8d}(a) and
our numerical scheme can accurately handle a wide range of gap sizes
thanks to the multi-level nature of the RCIP
method. Figure~\ref{konfig14a} shows $\vert u^{\rm sc}(r)\vert$ and
estimated error in $u^{\rm sc}(r)$ for the four-strip waveguide with
$h$ increased to $h=0.9997$. We here obtain
$\sigma^{\rm sc}=2\cdot 10^{-20}$ at $b=7.3$, $k=0.4094240286992$, and
$d=1.918010486711$.

\begin{figure}[t]
  \centering
  \includegraphics[height=47mm]{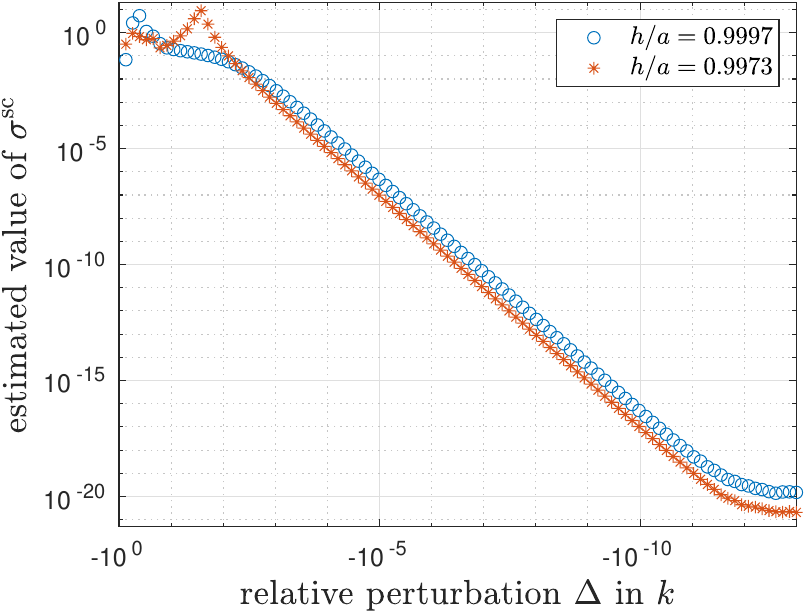}
  \includegraphics[height=47mm]{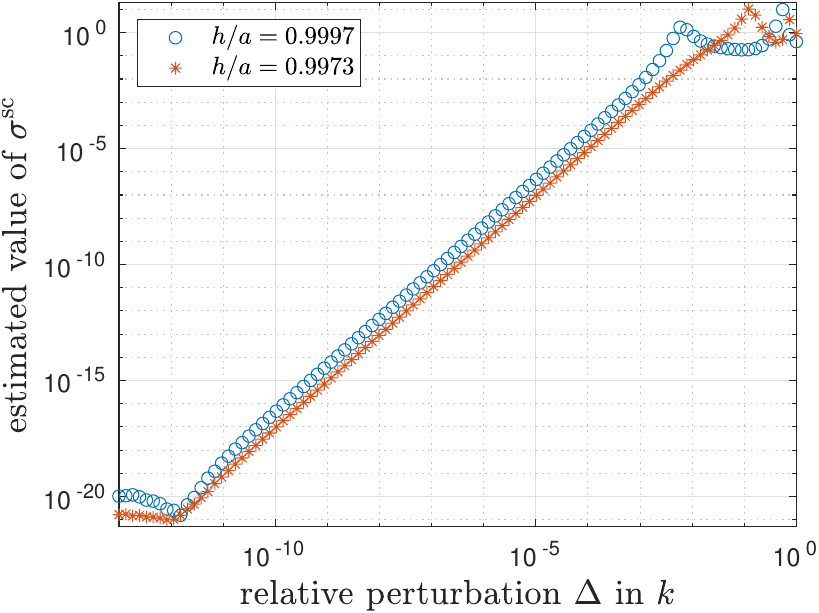}
  \caption{\sf $\sigma^{\rm sc}$ as a function of $\Delta$ for
    wavenumber $k(1+\Delta)$. Red asterisks refer to the waveguide in
    Figure~\ref{konfig8d} with $h=0.9973$ and
    $k=0.5712887729818$. Blue circles refer to the waveguide in
    Figure~\ref{konfig14a} with $h=0.9997$ and $k=0.4094240286992$.}
\label{bandwidth}
\end{figure}

\subsubsection{Bandwidth of four-strip waveguides}

The width of the frequency band where a B-L waveguide is invisible can
be adjusted, within certain limits, by changing $h$. This is
illustrated in Figure~\ref{bandwidth}, where the bandwidths of the
waveguides in Figure~\ref{konfig8d} and Figure~\ref{konfig14a} are
compared. The invisibility frequency bandwidth is tested by keeping
$b$, $d$, and $h$, changing $k$ to $k(1+\Delta)$ and evaluating
$\sigma^{\rm sc}$ in the interval $-1<\Delta\leq 1$.
Figure~\ref{bandwidth} shows that the relative bandwidth estimated at
$\sigma^{\rm sc}=10^{-10}$ is 2.20 times larger for $h=0.9973$ than
for $h=0.9997$, indicating that for a given number of strips, the
relative bandwidth decreases with increasing $h$. We also remark that
as $h\to a$, because of the corresponding smaller bandwidth, it will
be progressively more difficult to find a good initial guess for $k$
and $d$ in the optimization procedure.

\begin{figure}[t]
  \centering
  \includegraphics[height=47mm]{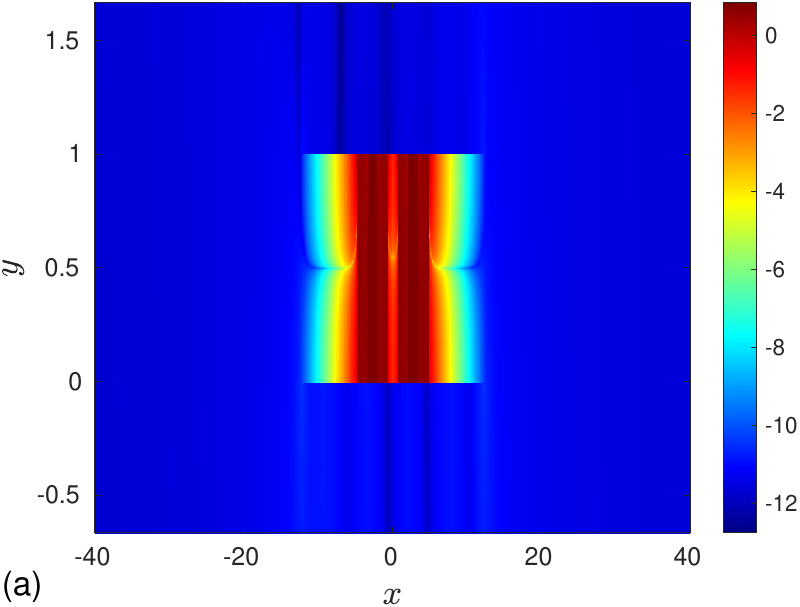}
  \includegraphics[height=47mm]{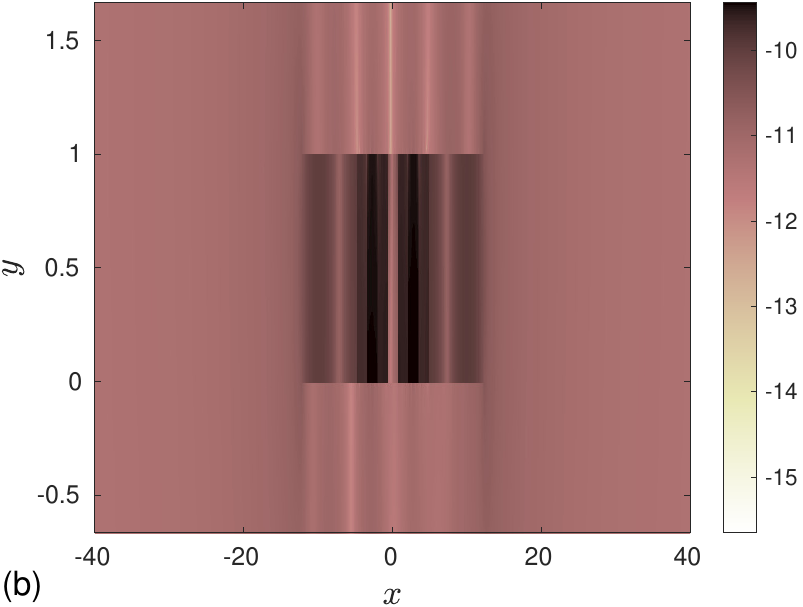}
  \caption{\sf The eight-strip waveguide: (a) $\log_{10}$ of
    $\vert u^{\rm sc}(r)\vert$; (b) $\log_{10}$ of estimated absolute
    pointwise error in $u^{\rm sc}(r)$.}
\label{konfig8d8p}
\end{figure}

\subsubsection{B-L waveguide with eight strips}

The eight-strip waveguide in Figure~\ref{konfig8d8p} has $b=7.3$ and
$h=0.9973$, just like the four-strip waveguide in
Figure~\ref{konfig8d}. Optimization gives
$\sigma^{\rm sc}=8\cdot 10^{-21}$ at $k=0.5710229705548$ and
$d=1.374587853864$. Thus $k$ and $d$ are almost the same as for the
four-strip waveguide. From \eqref{constraint2a}, \eqref{constraint2b},
and the field $\vert u^{\rm sc}(r)\vert$ in Figure~\ref{konfig8d8p}(a)
it follows that $u(r)$ of the eight-strip waveguide has $n=2$ and
$m=2$. This confirms that when the number of strips is doubled then
$n$ and $m$ are doubled while $k$ and $k_{\rm s}$ are essentially
unchanged.

\begin{figure}[t]
  \centering
  \includegraphics[height=47mm]{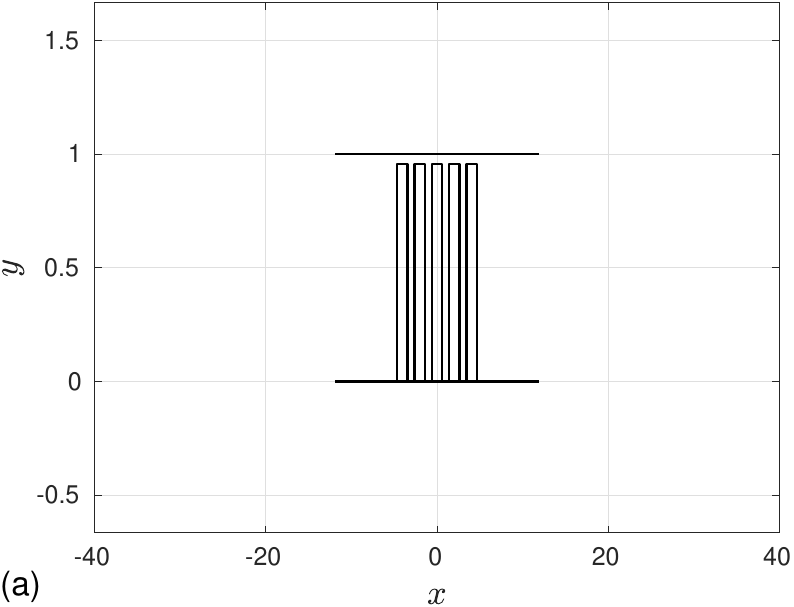}
  \includegraphics[height=47mm]{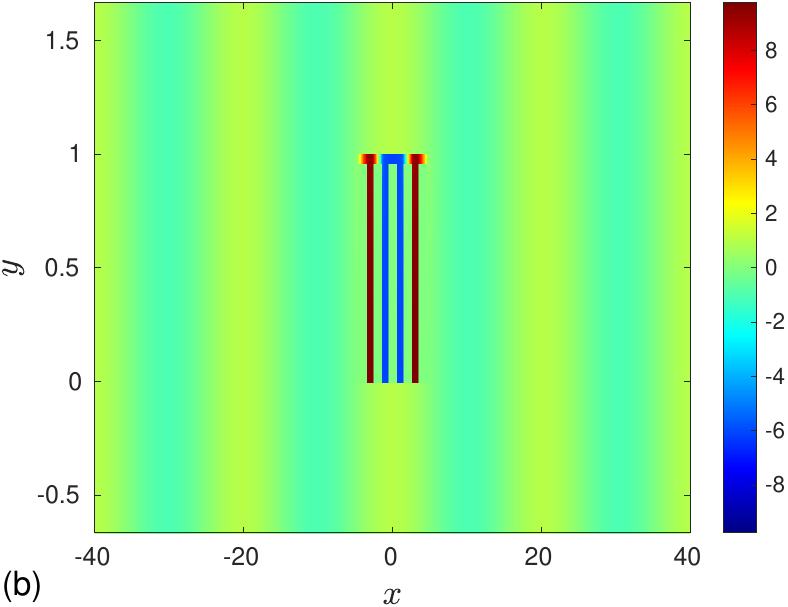}
  \includegraphics[height=47mm]{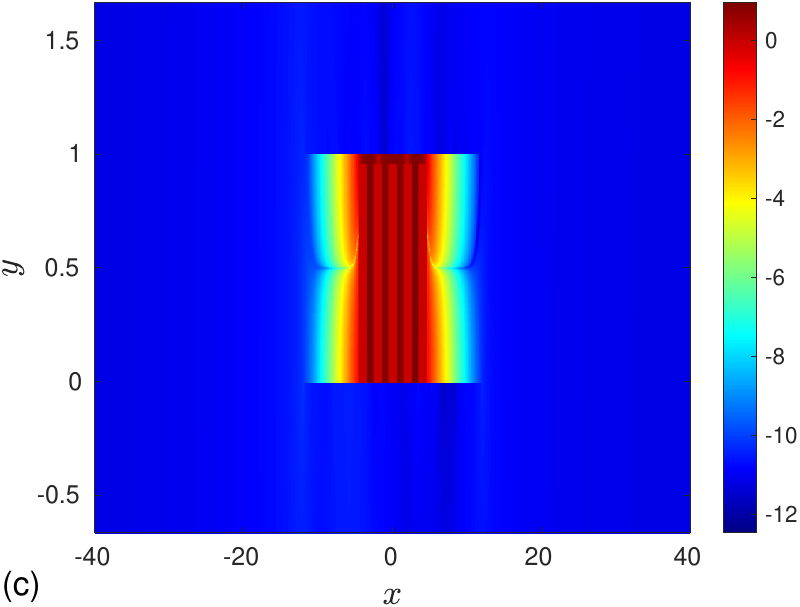}
  \includegraphics[height=47mm]{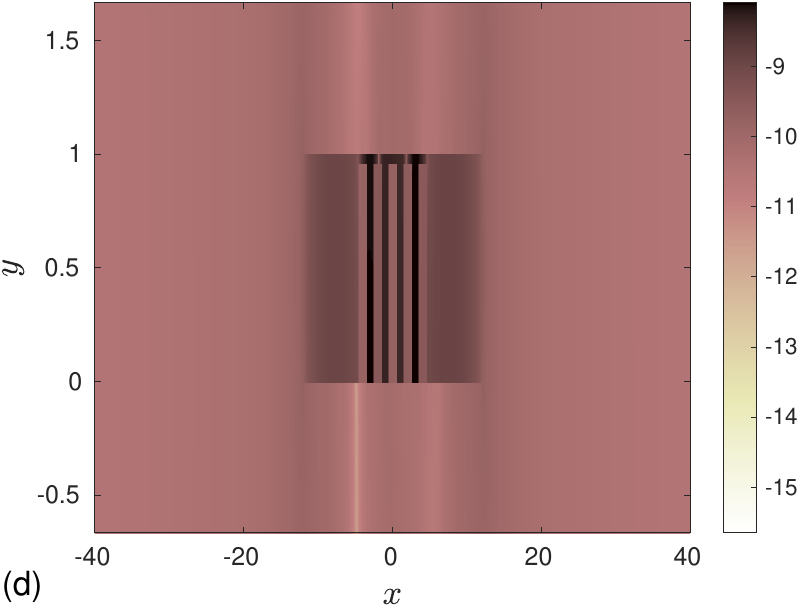}
  \caption{\sf (a) geometry of the five-bar waveguide; (b) real part
    of $u(r)$; (c) $\log_{10}$ of $\vert u^{\rm sc}(r)\vert$; (d)
    $\log_{10}$ of estimated absolute pointwise error in
    $u^{\rm sc}(r)$.}
\label{fivebar}
\end{figure}

\subsection{B-L waveguide with rectangular barriers}

The B-L waveguide in Figure~\ref{geometri2}(b) has five rectangular
bars and geometric parameters $a=1$, $b=7.3$, $h=0.957$, and
$w=1.2$. Optimization gives $\sigma^{\rm sc}\approx 4\cdot 10^{-19}$
at $k=0.3111846733919$ and $d=2.018021674709$. The corresponding
fields and error estimates are shown in Figure~\ref{fivebar}.

Figure~\ref{fivebar}(b) shows the evaluated real part of
$u(r)$. Inside the bars, where $u(r)$ should be zero, this field is
everywhere less than $10^{-11}$ in modulus, which is satisfactory.
Just like the four-strip waveguides in Figures~\ref{konfig8d} and
\ref{konfig14a}, the five-bar waveguide is seen to have wavenumbers
$k$ and $k_{\rm s}$ that satisfy~\eqref{constraint2a} and
\eqref{constraint2b} with $n=1$ and $m=1$. The errors in
Figure~\ref{fivebar}(d) are somewhat larger than in
Figures~\ref{konfig8d}(d), \ref{konfig14a}(b), and
\ref{konfig8d8p}(b), which may indicate that our numerical scheme is
slightly better at handling boundaries with many endpoints than
boundaries with many corners and branch points.

A bandwidth test, analogous to the one presented in
Figure~\ref{bandwidth}, was also made for the waveguide in
Figure~\ref{fivebar}. The resulting graph of $\sigma^{\rm sc}$ as a
function of $\Delta$ is almost indistinguishable from the red graph in
Figure~\ref{bandwidth} and therefore not shown.

\subsubsection{Resonances of inner regions}

The B-L waveguide in Figure~\ref{geometri2}(b) has a $\Gamma$ with
five closed boundary parts and it is appropriate here to comment on a
problem associated with such closed parts. The accuracy of numerical
solutions to BIEs modeling exterior problems may degrade for
wavenumbers close to those of inner resonances, as described in
\cite{Chew07}. However, the lowest resonance wavenumber for the
rectangles in Figure~\ref{fivebar}(a) is $\pi/w\approx 2.62$, which,
compared to $k=0.3111846733919$, is far too large to cause any
problems. Therefore, the problem of inner resonances is not an issue
in this work.

\subsection{Optimization in two steps}

To speed up the optimization, we first optimized the B-L waveguides
and corresponding wavenumbers using simplified geometries and the
Acoustic module in the FEM software package COMSOL Multiphysics
version 6.1. A simplified geometry is bounded and consists of a B-L
waveguide with an input port at the left end and an output port at the
right end. A TEM mode is launched at the input port. Based on given
values of geometric parameters, a wavenumber is identified for which
the reflected field at the input port is close to zero. By adjusting
the geometry, this wavenumber is shifted until the criterion
in~\eqref{constraint2a} is satisfied for $n=1$. The geometry and the
wavenumber are then further adjusted so that the reflected field and
the difference between the transmitted and incident fields are
minimized. The adjusted geometry and wavenumber so obtained then serve
as the initial guess for optimization using our BIE technique. As an
example, the initial guess for the waveguide in Figure~\ref{konfig8d}
is produced using the finest predefined mesh in COMSOL, with
$120,\!000$ degrees of freedom, and gives an initial-guess scattering
cross section of $\sigma^{\rm sc}\approx 7\cdot 10^{-4}$.

\subsection{Additional comments to the examples}

Our examples show that B-L waveguides can be rendered invisible for an
incident in-plane TM wave. In a narrow frequency band, the scattered
field is then close to zero everywhere outside the waveguide. This
means that several B-L waveguides can be combined to form a larger
invisible structure. Such a structure may be an alternative to an FSS
in applications that involve band pass structures. For the B-L
waveguide with rectangular bars, it is possible to hide other objects
inside the bars. This is then a perfect cloaking for a TM wave at
normal incidence and, for that wave, it is a much better cloaking than
that achieved by today's metamaterials. On the other hand, cloaking
with metamaterials has the advantage of being independent of the
polarization and angle of incidence of the incident plane wave.

The height $h$ of the barriers is close to $a$ (small gap) in all
examples. We have not tested for what values of $h$ invisible B-L
waveguides exist, but this is certainly an interesting issue. 

Another thing that should be interesting to investigate is the effect
of PEC-barrier shape on invisibility. That is, barriers of other
shapes than strips and bars. Even dielectric barriers might be an
option. Regardless of the type of barriers, it is crucial that the
barriers form a finite periodic array.

\section{Conclusions}
\label{sec:conclude}

We have shown that B-L waveguides can be designed to become invisible
in a frequency band of prescribed center frequency and, within certain
limits, width. The specific examples presented have barriers with
endpoints, corners, and branch points. We seek to design waveguides
with scattering cross sections on the order of $10^{-18}$. This
requires a robust numerical scheme that can accurately resolve the
Neumann Helmholtz problem close to singular boundary points. Such a
scheme has been developed by us and is reviewed and used in this work.

\begin{small}
\bibliographystyle{abbrv}
\bibliography{waveguide}

\begin{thebibliography}{10}

\bibitem{AgaOneRac23}
D.~Agarwal, M.~O'Neil, and M.~Rachh.
\newblock F{MM}-accelerated solvers for the {L}aplace-{B}eltrami problem on
  complex surfaces in three dimensions.
\newblock {\em J. Sci. Comput.}, 97(1):Paper No. 25, 25~pp., 2023.

\bibitem{Ande23}
E.~Andersson.
\newblock Implementation and study of boundary integral operators related to
  {PDE}:s in the plane.
\newblock Master's Theses in Mathematical Sciences 2023:E27, Centre for
  Mathematical Sciences, Lund University, 2023.
\newblock {\tt https://lup.lub.lu.se/student-papers/record/9133621}.

\bibitem{Anwar18}
R.~S. Anwar, L.~Mao, and H.~Ning.
\newblock Frequency selective surfaces: A review.
\newblock {\em Appl. Sci.}, 8(9):\;Paper No. 1689, 46~pp., 2018.

\bibitem{Arfken05}
G.~B. Arfken and H.~J. Weber.
\newblock {\em Mathematical methods for physicists}.
\newblock Elsevier, Amsterdam, sixth edition, 2005.

\bibitem{Bonnet17}
A.-S. Bonnet-Ben~Dhia, E.~Lun{\'e}ville, Y.~Mbeutcha, and S.~Nazarov.
\newblock A method to build non-scattering perturbations of two-dimensional
  acoustic waveguides.
\newblock {\em Math. Meth. Appl. Sci}, 40(2):335--349, 2017.

\bibitem{Boria07}
V.~E. Boria and B.~Gimeno.
\newblock Waveguide filters for satellites.
\newblock {\em IEEE Microw. Mag.}, 8(5):60--70, 2007.

\bibitem{BrunLint12}
O.~P. Bruno and S.~K. Lintner.
\newblock Second-kind integral solvers for {TE} and {TM} problems of
  diffraction by open arcs.
\newblock {\em Radio Sci.}, 47(6):\;Paper No. RS6006, 13~pp., 2012.

\bibitem{Chesnel22}
L.~Chesnel, J.~Heleine, and S.~A. Nazarov.
\newblock Acoustic passive cloaking using thin outer resonators.
\newblock {\em Z. Angew. Math. Phys.}, 73(3):\;Paper No. 98, 31~pp., 2022.

\bibitem{Chew07}
W.~Chew and J.~Song.
\newblock Gedanken experiments to understand the internal resonance problems of
  electromagnetic scattering.
\newblock {\em Electromagn.}, 27(8):457--471, 2007.

\bibitem{ColtonKress98}
D.~Colton and R.~Kress.
\newblock {\em Inverse acoustic and electromagnetic scattering theory},
  volume~93 of {\em Applied Mathematical Sciences}.
\newblock Springer-Verlag, Berlin, second edition, 1998.

\bibitem{Fan21}
J.~Fan, L.~Zhang, S.~Wei, Z.~Zhang, S.-K. Choi, B.~Song, and Y.~Shi.
\newblock A review of additive manufacturing of metamaterials and developing
  trends.
\newblock {\em Mater. Today}, 50:303--328, 2021.

\bibitem{Gbur13}
G.~Gbur.
\newblock Invisibility physics: past, present, and future.
\newblock In {\em Progress in Optics}, volume~58, pages 65--114. Elsevier,
  2013.

\bibitem{fmm2d}
Z.~Gimbutas, L.~Greengard, M.~O'Neil, M.~Rachh, and V.~Rokhlin.
\newblock {\it fmm2d software library}.
\newblock {\tt https://github.com/flatironinstitute/fmm2d}.
\newblock Accessed: 2023-04-17.

\bibitem{HaBaMaYo14}
S.~Hao, A.~H. Barnett, P.~G. Martinsson, and P.~Young.
\newblock High-order accurate methods for {N}ystr\"{o}m discretization of
  integral equations on smooth curves in the plane.
\newblock {\em Adv. Comput. Math.}, 40(1):245--272, 2014.

\bibitem{Tutorial}
J.~Helsing.
\newblock Solving integral equations on piecewise smooth boundaries using the
  {RCIP} method: a tutorial.
\newblock {\em arXiv e-prints}, arXiv:1207.6737v10 [physics.comp-ph], revised
  2022.

\bibitem{HelsJian23}
J.~Helsing and S.~Jiang.
\newblock Solving {D}irichlet and {N}eumann {H}elmholtz problems on piecewise
  smooth open curves.
\newblock 2023.
\newblock In preparation.

\bibitem{HelsKarl13}
J.~Helsing and A.~Karlsson.
\newblock An accurate boundary value problem solver applied to scattering from
  cylinders with corners.
\newblock {\em IEEE Trans. Antennas Propag.}, 61(7):3693--3700, 2013.

\bibitem{HelsKarl18}
J.~Helsing and A.~Karlsson.
\newblock On a {H}elmholtz transmission problem in planar domains with corners.
\newblock {\em J. Comput. Phys.}, 371:315--332, 2018.

\bibitem{HelsOjal09}
J.~Helsing and R.~Ojala.
\newblock Elastostatic computations on aggregates of grains with sharp
  interfaces, corners, and triple-junctions.
\newblock {\em Internat. J. Solids Structures}, 46(25):4437--4450, 2009.

\bibitem{Kadic19}
M.~Kadic, G.~W. Milton, M.~van Hecke, and M.~Wegener.
\newblock 3{D} metamaterials.
\newblock {\em Nat. Rev. Phys.}, 1(3):198--210, 2019.

\bibitem{ShipTu12}
S.~P. Shipman and H.~Tu.
\newblock Total resonant transmission and reflection by periodic structures.
\newblock {\em SIAM J. Appl. Math.}, 72(1):216--239, 2012.

\end{thebibliography}
\end{small}

\end{document}